\documentclass[12pt]{aastex}
\usepackage{epsfig,amsmath}
\newcommand{\casa}{Cassiopeia A }

\shortauthors{VERITAS Collaboration}

\shorttitle{Observations of Cassiopeia A with VERITAS}

\begin{document}

\title{Observations of the shell-type SNR Cassiopeia A at TeV energies with VERITAS}

\author{
V.~A.~Acciari\altaffilmark{1},
E.~Aliu\altaffilmark{2},
T.~Arlen\altaffilmark{3},
T.~Aune\altaffilmark{4},
M.~Bautista\altaffilmark{5},
M.~Beilicke\altaffilmark{6},
W.~Benbow\altaffilmark{1},
D.~Boltuch\altaffilmark{2},
S.~M.~Bradbury\altaffilmark{7},
J.~H.~Buckley\altaffilmark{6},
V.~Bugaev\altaffilmark{6},
Y.~Butt\altaffilmark{8},
K.~Byrum\altaffilmark{9},
A.~Cannon\altaffilmark{10},
A.~Cesarini\altaffilmark{11},
Y.~C.~Chow\altaffilmark{3},
L.~Ciupik\altaffilmark{12},
P.~Cogan\altaffilmark{5},
W.~Cui\altaffilmark{13},
R.~Dickherber\altaffilmark{6},
C.~Duke\altaffilmark{14},
T.~Ergin\altaffilmark{8},
S.~J.~Fegan\altaffilmark{3,\dag},
J.~P.~Finley\altaffilmark{13},
G.~Finnegan\altaffilmark{15},
P.~Fortin\altaffilmark{16,\dag},
L.~Fortson\altaffilmark{12},
A.~Furniss\altaffilmark{4},
N.~Galante\altaffilmark{1},
D.~Gall\altaffilmark{13},
G.~H.~Gillanders\altaffilmark{11},
J.~Grube\altaffilmark{10},
R.~Guenette\altaffilmark{5},
G.~Gyuk\altaffilmark{12},
D.~Hanna\altaffilmark{5},
J.~Holder\altaffilmark{2},
D.~Huang\altaffilmark{17},
C.~M.~Hui\altaffilmark{15},
T.~B.~Humensky\altaffilmark{18},
P.~Kaaret\altaffilmark{19},
N.~Karlsson\altaffilmark{12},
M.~Kertzman\altaffilmark{20},
D.~Kieda\altaffilmark{15},
A.~Konopelko\altaffilmark{17,*},
H.~Krawczynski\altaffilmark{6},
F.~Krennrich\altaffilmark{21},
M.~J.~Lang\altaffilmark{11},
S.~LeBohec\altaffilmark{15},
G.~Maier\altaffilmark{5,\%},
S.~McArthur\altaffilmark{6},
A.~McCann\altaffilmark{5},
M.~McCutcheon\altaffilmark{5},
J.~Millis\altaffilmark{13,\$},
P.~Moriarty\altaffilmark{22},
R.~A.~Ong\altaffilmark{3},
D.~Pandel\altaffilmark{19},
J.~S.~Perkins\altaffilmark{1},
M.~Pohl\altaffilmark{21,\&},
J.~Quinn\altaffilmark{10},
K.~Ragan\altaffilmark{5},
P.~T.~Reynolds\altaffilmark{23},
E.~Roache\altaffilmark{1},
H.~J.~Rose\altaffilmark{7},
M.~Schroedter\altaffilmark{21},
G.~H.~Sembroski\altaffilmark{13},
A.~W.~Smith\altaffilmark{9},
B.~R.~Smith\altaffilmark{17},
D.~Steele\altaffilmark{12,\#},
S.~P.~Swordy\altaffilmark{18},
M.~Theiling\altaffilmark{1},
S.~Thibadeau\altaffilmark{6},
A.~Varlotta\altaffilmark{13},
V.~V.~Vassiliev\altaffilmark{3},
S.~Vincent\altaffilmark{15},
R.~G.~Wagner\altaffilmark{9},
S.~P.~Wakely\altaffilmark{18},
J.~E.~Ward\altaffilmark{10},
T.~C.~Weekes\altaffilmark{1},
A.~Weinstein\altaffilmark{3},
T.~Weisgarber\altaffilmark{18},
S.~Wissel\altaffilmark{18},
M.~Wood\altaffilmark{3}
}

\altaffiltext{1}{Fred Lawrence Whipple Observatory, Harvard-Smithsonian Center for Astrophysics, Amado, AZ 85645, USA}
\altaffiltext{2}{Department of Physics and Astronomy and the Bartol Research Institute, University of Delaware, Newark, DE 19716, USA}
\altaffiltext{3}{Department of Physics and Astronomy, University of California, Los Angeles, CA 90095, USA}
\altaffiltext{4}{Santa Cruz Institute for Particle Physics and Department of Physics, University of California, Santa Cruz, CA 95064, USA}
\altaffiltext{5}{Physics Department, McGill University, Montreal, QC H3A 2T8, Canada}
\altaffiltext{6}{Department of Physics, Washington University, St. Louis, MO 63130, USA}
\altaffiltext{7}{School of Physics and Astronomy, University of Leeds, Leeds, LS2 9JT, UK}
\altaffiltext{8}{Harvard-Smithsonian Center for Astrophysics, 60 Garden Street, Cambridge, MA 02138, USA}
\altaffiltext{9}{Argonne National Laboratory, 9700 S. Cass Avenue, Argonne, IL 60439, USA}
\altaffiltext{10}{School of Physics, University College Dublin, Belfield, Dublin 4, Ireland}
\altaffiltext{11}{School of Physics, National University of Ireland, Galway, Ireland}
\altaffiltext{12}{Astronomy Department, Adler Planetarium and Astronomy Museum, Chicago, IL 60605, USA}
\altaffiltext{13}{Department of Physics, Purdue University, West Lafayette, IN 47907, USA }
\altaffiltext{14}{Department of Physics, Grinnell College, Grinnell, IA 50112-1690, USA}
\altaffiltext{15}{Department of Physics and Astronomy, University of Utah, Salt Lake City, UT 84112, USA}
\altaffiltext{16}{Department of Physics and Astronomy, Barnard College, Columbia University, NY 10027, USA}
\altaffiltext{17}{Department of Physics, Pittsburg State University, 1701 South Broadway, Pittsburg, KS 66762, USA}
\altaffiltext{18}{Enrico Fermi Institute, University of Chicago, Chicago, IL 60637, USA}
\altaffiltext{19}{Department of Physics and Astronomy, University of Iowa, Van Allen Hall, Iowa City, IA 52242, USA}
\altaffiltext{20}{Department of Physics and Astronomy, DePauw University, Greencastle, IN 46135-0037, USA}
\altaffiltext{21}{Department of Physics and Astronomy, Iowa State University, Ames, IA 50011, USA}
\altaffiltext{22}{Department of Life and Physical Sciences, Galway-Mayo Institute of Technology, Dublin Road, Galway, Ireland}
\altaffiltext{23}{Department of Applied Physics and Instrumentation, Cork Institute of Technology, Bishopstown, Cork, Ireland}
\altaffiltext{\$}{Now at Department of Physics, Anderson University, 1100 East 5th Street, Anderson, IN 46012}
\altaffiltext{\&}{Now at Institut f\"ur Physik und Astronomie, Universit\"at Potsdam, 14476 Potsdam-Golm, Germany;
DESY, Platanenallee 6, 15738 Zeuthen, Germany}
\altaffiltext{\%}{Now at DESY, Platanenallee 6, 15738 Zeuthen, Germany}
\altaffiltext{\#}{Now at Los Alamos National Laboratory, MS H803, Los Alamos, NM 87545}
\altaffiltext{\dag}{Now at Laboratoire Leprince Ringuet, Ecole Polytechnique, CNRS, IN2P3, Palaiseau, France}
\altaffiltext{*}{corresponding author}

\begin{abstract}
We report on observations of very high-energy $\gamma$ rays from the shell-type
supernova remnant Cassiopeia~A with the VERITAS stereoscopic array of four
imaging atmospheric Cherenkov telescopes in Arizona. The total exposure time
for these observations is 22~hours, accumulated between September and
November of 2007. The $\gamma$-ray source associated with the SNR Cassiopeia~A
was detected above 200~GeV with a statistical significance of
8.3\,$\sigma$. The estimated integral flux for this $\gamma$-ray source is about 3\%
of the Crab-Nebula flux. The photon spectrum is compatible with a power law
$dN/dE \propto E^{-\Gamma}$ with an index $\Gamma=2.61 \pm 0.24_{stat} \pm
0.2_{sys}$. The data are consistent with a point-like source. 
We provide a detailed description of the analysis results, and discuss
physical mechanisms that may be responsible for the observed
$\gamma$-ray emission.
\end{abstract}

\section{Introduction}
Cassiopeia~A is the youngest of the 
%known 
historical galactic supernova remnants (SNR); it first appeared 
in the sky about 1680 \citep{Ashworth1980,StephensonGreen2002}. 
The best estimate of the actual Cassiopeia A SN explosion date is AD 1680.5$\pm$18.7,  
which was deduced from the HST ({\it Hubble Space Telescope}) measurements of the expansion of ejecta knots
\citep{Fesen06}.
It is also the brightest and one of the best studied
radio sources in the sky (e.g, \citet{Kassimetal.1995}). Located 3.4~kpc away
\citep{Reedetal.1995}, the optical shell of $2.5\arcmin$ radius corresponds to a physical size of about 2.5~pc.
The synchrotron radiation of Cassiopeia~A extends from radio wavelengths through the sub-millimeter
\citep{Mezgeretal.1986} and near-infrared \citep{Tuffsetal.1997} all the way to hard X-rays
\citep{Allenetal.1997,Favataetal.1997,Vink01,Renaud06}. 
The nature of the hard X-ray emission measured up to 100~keV still remains unclear (see \cite{HelderVink08}), albeit 
the morphology of the non-thermal X-ray emission is dominated by
faint, well-defined filaments and knots \citep{Hughesetal.2000,VinkLaming2003}, which are possibly sites 
of cosmic ray acceleration. These energetic hadronic cosmic rays (CR) can subsequently produce $\gamma$ 
rays in collisions with ambient gas via  $\pi^\circ$-decay \citep{Druryetal1994}.

EGRET did not detect $\gamma$-ray emission above 100~MeV from Cassiopeia~A
\citep{Espositoetal.1996}. Likewise, first attempts to detect TeV $\gamma$-ray emission from the ground with 
the 10~m Whipple telescope \citep{Lessardetal1999} and with CAT \citep{Goretetal1999} resulted only in 
upper limits. With an exposure of 232~hours, accumulated
during the summer months of 1997, 1998, and 1999, HEGRA detected TeV $\gamma$-ray
emission associated with Cassiopeia~A \citep{HEGRA2000}. A 5\,$\sigma$ detection of Cassiopeia~A
resulted in a flux estimate of $(5.8\pm1.2_{stat}\pm 1.2_{syst}) \times 10^{-13} \, \rm cm^{-2} s^{-1}$ above 1~TeV.
The energy spectrum measured in a range from 1 to 10~TeV was consistent with a power law with a photon
index of $\Gamma = 2.5\pm0.4_{stat} \pm0.1_{syst}$. Recently MAGIC and VERITAS have made observations of Cassiopeia~A at a substantially lower
energy threshold.

Cassiopeia~A was observed with the 17~m MAGIC telescope between July 2006 and January 2007 for a total
exposure of 47~hours \citep{Albertetal2007}. The $\gamma$-ray source was detected above 250~GeV at the
level of statistical significance of 5.2\,$\sigma$, with a photon flux above 1 TeV of
$(7.3 \pm 0.7_{stat} \pm 2.2_{syst})\times 10^{-13}\, \rm cm^{-2}s^{-1}$. The photon spectrum is compatible
with a power law with an index $\Gamma = 2.3\pm 0.2_{stat}\pm 0.2_{syst}$. The source is point-like
for the given angular resolution of the telescope. The position of the MAGIC source is consistent with the source
position previously published by HEGRA. In addition, the energy spectrum measured by MAGIC agrees within
statistical errors with that measured by HEGRA.

Here we report on observations of Cassiopeia~A with the VERITAS stereoscopic array of four imaging atmospheric
Cherenkov telescopes in Arizona.  In this paper a short description of the experiment is followed by
a summary of observational data and analysis results. Finally a discussion of the physics
implications of VERITAS data for existing models of TeV $\gamma$-ray emission from Cassiopeia~A is given.

\section{Experiment}

VERITAS ({\em Very Energetic Radiation Imaging Telescope Array System}) \citep{Weekes2002} is an array
of four imaging atmospheric Cherenkov telescopes located in southern Arizona ($31^\circ40\arcmin$ N,
$110^\circ 57\arcmin$ W) at an altitude of 1.3~km. The four VERITAS telescopes are almost identical
in their technical parameters \citep{Holderetal2006}. The 12~m optical reflector of a VERITAS telescope is a
tessellated structure consisting of 357 identical spherical mirror facets, which are hexagonal in shape. The arrangement of the mirror facets constitutes a Davies-Cotton design \citep{DaviesCotton1957}, providing a total reflecting area of 110~$\rm m^2$.
The point-spread function of a VERITAS telescope has a FWHM of $\sim4\arcmin$ on-axis \citep{McCann}. A high-resolution imaging
camera placed at the focus of the reflector consists of 499 photomultiplier tubes (PMTs) in a close-packed hexagonal
arrangement and has a field of view of $3.5^\circ$. Each camera PMT views a circle of diameter $0.15^\circ$ on the sky. A set of light concentrators is mounted in front of the PMTs to increase the light-collection efficiency and block the off-axis 
light. The camera triggers if the signal in each of any three adjacent PMTs exceeds a discriminator threshold
of 50~mV, corresponding to  approximately 4-5 photoelectrons. A coincidence of at least two cameras triggering within
a time gate of 100~ns is required to read out an event. A 48 ns (24-samples) length of each PMT signal is digitized with custom-built 500~MegaSamples/s flash-ADC electronics. The nominal trigger rate of the
four-telescope array was about 230~Hz at zenith. The cameras are flat-fielded and calibrated using nightly measured
laser runs. The pedestal and pedestal variances ($\sigma$), which provide a measure of the night sky background
noise level, were calculated during each data run using pedestal events, injected at a 1~Hz frequency.
The pedestal variances were used for computing the dynamic {\em picture}  and {\em boundary}  thresholds for consequent
image cleaning. All PMTs with a signal exceeding the {\it picture} threshold (5$\sigma$) were used in the image parametrization. 
PMTs with a signal exceeding the {\it boundary} threshold (2.5$\sigma$) but lying near the {\it picture} PMTs were also selected. 
(e.g., see \citet{Holderetal2006}). To characterize the shape and orientation of calibrated images
recorded by each telescope, the standard second-moment parameters were calculated as described by
\citet{rey93}.

\section{Summary of Data}

Cassiopeia~A was observed with VERITAS for 22~hours between September and November of
2007. All observations were made with the full four-telescope array during moonless nights.
The data-analysis pipeline consists of two distinct phases. After the data are processed, 
the distributions from the raw data are accumulated as diagnostics of both the
instrument performance and the stability of the weather conditions. Each data run is
inspected for rate, timing and tracking consistency, and either accepted or rejected based
on this first pass. Once this diagnostic pass is made, acceptable runs are further analyzed.
All data were taken in 20-minute runs using the so-called {\em Wobble}
source-tracking mode, which is optimal for observations of a point-like source. In {\em Wobble}
mode the source is positioned at a $0.5^\circ$ offset from the center of the field of view of the
camera during observations, which allows for both on-source observations and simultaneous
estimation of the background contamination caused by charged cosmic rays. The number
of background events in the signal region was estimated using a number of regions distributed symmetrically
with respect to the center of the camera for each wobble offset. The offset directions towards North,
South, East, or West, were consequently alternated on a run-by-run basis. A total of 74 data runs
were collected at zenith angles between $26^\circ$ and $39^\circ$. The average zenith angles and the
average event trigger rate were $31.5^\circ \pm3.7^\circ$ and $232\pm12$~Hz, respectively.
Prior to applying analysis cuts, data were selected for adequate image quality, by requiring
a minimum integrated charge of all pixels in the image  of 400~digital counts
(approximately 80 photo-electrons) and a maximum distance of the image's centroid from the center of the
field of view  of 1.43$^\circ$. These cuts were {\it a priori} optimized using the Crab Nebula data sample. 
Each accepted event was also required to contain at least two images passing these cuts. 
The VERITAS experimental setup during Cassiopeia~A observations 
included two telescopes placed at a rather small separation of 35~m. 
%It appears that the two-fold
Coincidence events including both of these telescopes 
%are extremely disadvantageous for adequate
%stereoscopic reconstruction. Therefore, such events 
have been removed from the analysis. 
During the summer of 2009, one of these telescopes was relocated in order to improve the sensitivity of the array.

\section{Data Analysis}

The imaging analysis of the VERITAS data is based on the reconstruction of the shower direction
for each individual event \citep{Konopelkoetal1999,Hofmannetal1999,Krawczynskietal2006}, and
joint parametrization of the shape of the Cherenkov light flash from an individual shower using
a multiple-telescope approach \citep{Konopelko1995,Krawczynskietal2006}. All recorded
events were subjected to the canonical directional cut on  $\theta^2$, where $\theta$ is the angular
distance between the true source position on the sky and the reconstructed one. Of the remaining events,
the candidates for $\gamma$-ray showers were selected using two simultaneously
applied cuts on the parameters of image shape: MSW ({\em mean scaled Width}) and MSL ({\em mean
scaled Length}). These three major analysis cuts were optimized using Crab Nebula observational
data from the same epoch, chosen for the same zenith angle range as covered in observations
of Cassiopeia~A. The choice of optimal analysis cuts depends noticeably on
the flux of the putative $\gamma$-ray source. Therefore, we developed two sets of
analysis cuts, appropriate for flux levels of 1 and 0.03 Crab (see Table~\ref{t1}) respectively. Both sets of  optimal
analysis cuts yield comparable signal significances for the Crab Nebula as well as Cassiopeia~A (Table~\ref{t2}).

\begin{figure}[btp]
\includegraphics [width=1.0\textwidth]{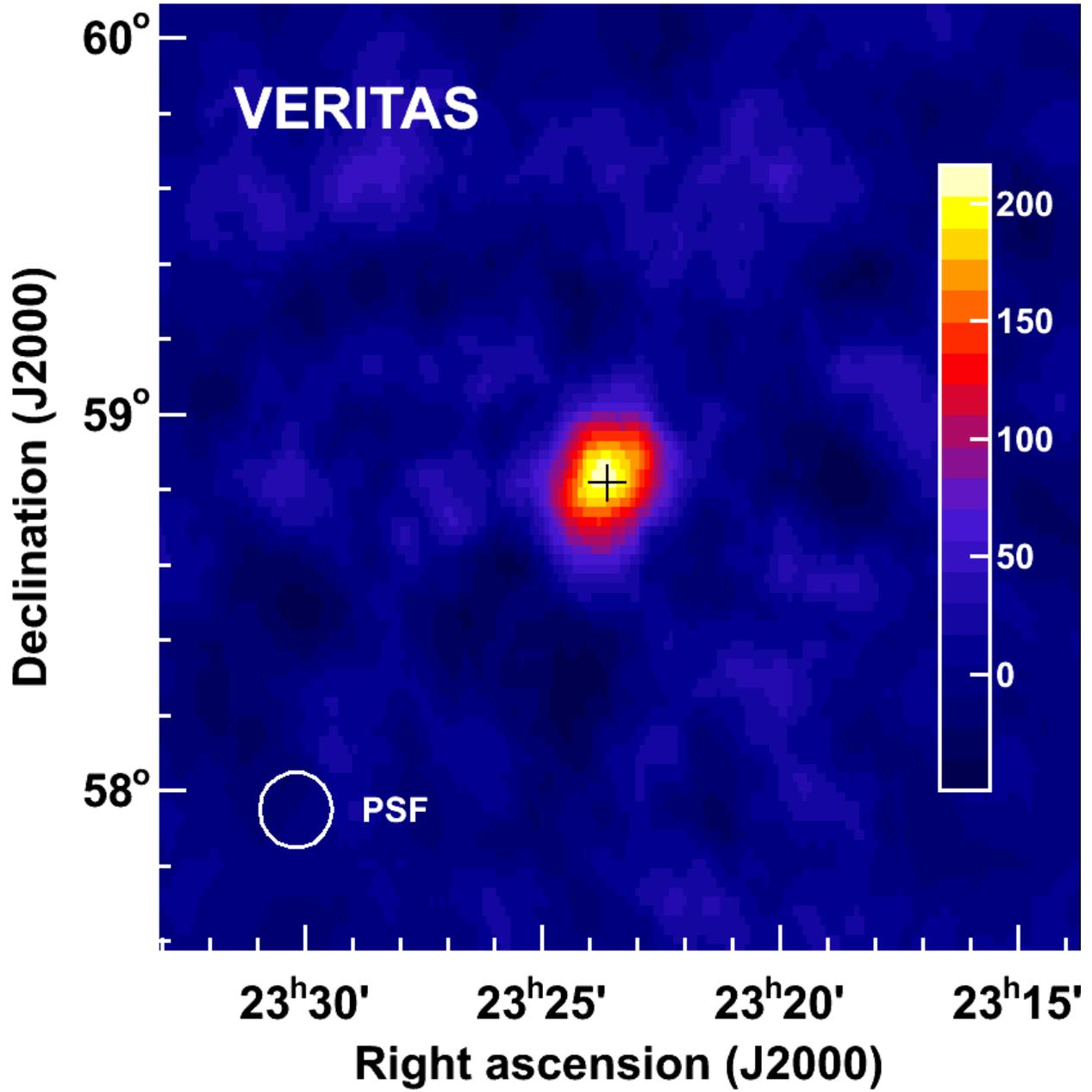}
\caption{Smoothed sky map of excess
counts from the region centered at Cassiopeia~A observed with VERITAS for a total of 22 hours in 2007. The color bar
represents the excess event counts. The white circle indicates the size of the VERITAS point-spread function. The cross indicates the measured position of the TeV $\gamma$-ray source. The radius  of a smoothing circular window was $0.115^\circ$.}
\label{fg2}
\end{figure}

The VERITAS array enables measurement of
the arrival direction of every individual shower detected. All recorded events that have passed both the image
quality cuts and specific analysis cuts can be plotted in a two-dimensional sky map.
Even after applying rather strict selection criteria such maps are dominated by the
flux of the isotropic cosmic ray background. 
%To extract $\gamma$-ray emission from the background is a major challenge
%for ground-based Cherenkov telescopes. 
A number of methods (background models) have been developed for effective removal
of background \citep{berge2006}. These models can effectively handle the background
issues of diverse observations, but the weaknesses or strengths of any particular approach depend
on the flux, angular morphology, and spectrum of a given $\gamma$-ray source. In this paper we used a method that is
rather stable with respect to any systematic background inhomogeneity across the camera field of view, the so-called
ring-background model. In this model, a ring (annulus) around the location of a putative $\gamma$-ray source in the camera
focal plane provides an immediate background estimate. The canonical angular radius of the background ring is $0.5^\circ$,
whereas the angular area (solid angle) covered by the ring is typically chosen to be larger than
that of the circular source region by a factor of 7-10. The excess map of the sky region around Cassiopeia~A for the data
set of 22~hours is shown in Figure~\ref{fg2}. An evident excess due to $\gamma$ rays at a 8.3\,$\sigma$ level of statistical
significance \citep{LiMa83} can be observed at the position of Cassiopeia~A. This result has been cross-checked using a standard {\it Wobble} analysis. 
%The results of the analysis using the ring-background model are summarized in
%Table~\ref{t2}.

\section{Source Localization}

The energy-averaged angular resolution of the VERITAS array for an individual $\gamma$-ray event is approximately $4\arcmin$ to $6\arcmin$ (68\% containment radius). This implies that a point-like $\gamma$-ray source detected by VERITAS appears as a spot of finite size in the expanded $\gamma$-ray sky map. The centroid of this spot is taken as the coordinates of the putative $\gamma$-ray emitter. Any error in the telescope pointing direction will deteriorate the
exact measurement of the $\gamma$-ray source position. Note that the pointing accuracy of the telescopes is limited,
by small misalignment of azimuth and altitude axes, and elastic deformations of the telescope structure. These effects contribute to the mispointing of the array, which strongly depends on the altitude and azimuth of observation. Most of the
pointing uncertainties can be substantially diminished by taking pointing calibration data on a monthly basis. Each
telescope is pointed at a number of bright stars uniformly distributed on the sky. The star is imaged by the telescope
mirror onto a screen in front of the Cherenkov camera, and the image is recorded by a CCD camera. The position
of each spot is then compared to the nominal center of the Cherenkov camera. These results contribute to a multi-parameter
pointing model in the telescope tracking software that corrects for the measured misalignment during observation.
This procedure was extensively tested on a number of VHE $\gamma$-ray point sources of known position. In addition, the residual mispointing can be evaluated from a detailed comparison of the nominal position of the source evaluated for the
different wobble offsets, energy thresholds, telescope multiplicities, analysis cuts, etc. During Cassiopeia~A observations
the systematic pointing error of the VERITAS array is $\sim 1.2\arcmin$.

The measured position of the $\gamma$-ray source is determined by a fit over a circular window of $0.5^\circ$ radius
centered on Cassiopeia~A using its known coordinates. The profile of the $\gamma$-ray excess can be modeled by the
two-dimensional Gaussian distribution:
\begin{equation}
f(\theta_x, \theta_y) \propto exp(- \frac{1}{2} ( \frac{ (\theta_x- {\theta_x}_o)^2 }{ {\sigma^2}_x } +
\frac{(\theta_y - {\theta_y}_o)^2 } {{\sigma^2}_y}),
\label{fit}
\end{equation}
where $ {\theta_x}_o, {\theta_y}_o$ are the angular coordinates of the $\gamma$-ray emission centroid, and ${\sigma^2}_x , {\sigma^2}_y$ are the extensions of the signal region in two perpendicular directions.
The width of the two-dimensional Gaussian fit is composed of the fixed angular resolution of the VERITAS array and the
intrinsic size of the source. Note that the excess map generated by the ring-background model has been smoothed using
a circular window of $0.115^\circ$ radius, which approximately corresponds to the angular resolution of the VERITAS
array.
%To gain even further in angular resolution, and also remove any systematic biases in stereoscopic reconstruction,
%the reduced data, containing  only 3- and 4-fold telescope events, have to pass an additional requirement of a higher size
%cut (650 digital counts) used for the source localization.
%The four-dimensional minimization procedure was exploited to obtain the
%best-fit position. 
First, this method was tested on the 1ES~2344+514 data taken with the VERITAS telescope array during the
same observational season as that of \casa with an instrument of similar configuration. 1ES~2344+514 is a blazar-type AGN,
which was in a high state of $\gamma$-ray emission during the Cassiopeia~A observations. Given the redshift of 1ES~2344+514 of z = 0.044, it is indubitably a point source. In addition, the total number of recorded $\gamma$ rays from 1ES~2344+514 was of the same order as the number of excess counts from Cassiopeia~A. Therefore, this object could be used as a calibration source for estimating the limits of the source localization procedure.  The width of the $\gamma$-ray point-spread function (PSF) evaluated using 1ES~2344+514 data is $\sigma_o = 4.8\arcmin$. However, this observationally
determined PSF is significantly affected by the angular size of the signal region used by the ring-background model,
which was adopted for the smoothing of the two-dimensional sky maps. Alternatively, one can use the excess count
sky map of uncorrelated bins which leads to similar results. The position of the 1ES~2344+514 $\gamma$-ray peak derived from the best fit was found to be consistent with the astronomical position of this object (RA ($\alpha$): 23 h 47 m 04.919 s, dec ($\delta$): $\rm +51^o\, 42\arcmin \, 17.87\arcsec$) within the statistical uncertainties of the best-fit position on right ascension and declination, $\Delta_{RA}$ and $\Delta_{dec}$, of $1.24\arcmin$ and
$9\arcsec$, respectively. These results ultimately validate the accuracy of the analysis method.
Cassiopeia~A data were analyzed using exactly the same two-dimensional analysis technique. The derived position of
the peak of $\gamma$-ray emission from Cassiopeia~A deviates from the nominal position of the supernova remnant \citep{Becker91}
($\alpha$ = 23 h 23 m 24 s, $\delta$ = $\rm +58\deg\, 48.9\arcmin$) by less than $\Delta_{RA}=14\arcsec$ and $\Delta_{dec}=35\arcsec$.
Evidently, the observed $\gamma$-ray emission is associated with the Cassiopeia~A SNR.

\section{Source Extension}

The angular radius of Cassiopeia~A, measured at wavelengths longer than those corresponding
to TeV energies, is about 2.5$\arcmin$. Primarily, this can be used as a characteristic angular size of 
the TeV $\gamma$-ray source. It is apparent that such a small angular dimension of Cassiopeia~A 
is well below the angular resolution (PSF) of VERITAS, $\sigma_\circ \simeq 4.8 \arcmin$, which 
unavoidably smears out the intrinsic source distribution and consequently does not permit detailed 
mapping of the morphology of the $\gamma$-ray source. The angular profile of the observed $\gamma$-ray 
peak finally constrains the intrinsic angular size of the source. The two-dimensional, azimuthally symmetric 
Gaussian function could be naturally 
used to model the measured angular shape of the $\gamma$-ray signal. The angular extent of the $\gamma$-ray 
peak towards Cassiopeia A measured with VERITAS is $\sigma_{Cas\, A}=$5.3$\pm$0.5$\arcmin$.  

The angular extent of the point-spread function can be derived from the data taken on a calibration $\gamma$-ray
source. For that we can use again the contemporaneous observations of 1ES~2344+514. The best fit of the $\gamma$-ray 
peak for 1ES~2344+514 gives $\sigma$ = 4.8$\pm$0.2$\arcmin$. 
If we assume a $\gamma$-ray source with a Gaussian profile, an approximate upper limit on the source extent 
can be calculated by summing the measured extents of the PSF and the Cassiopeia A $\gamma$-ray signal in quadrature, 
$\sigma_{s}=(\sigma_{Cas\,A}^2-\sigma^2)^{1/2}\lesssim 3.5\arcmin$. Thus, given rather large 
statistical errors of the involved angular extents the shape of the Cassiopeia~A 
signal is hardly distinguishable from the point-spread function and the $\gamma$-ray signal is statistically consistent with the point 
source.  

The low statistics of currently recorded $\gamma$ rays from Cassiopeia~A are not sufficient to draw a final conclusion 
on the source extension.
% and any upper limit on its angular extent would not provide any scientifically interesting constraints.
A further, deeper, observation of Cassiopeia~A with VERITAS might help to improve the measurement of the angular extension of 
the TeV $\gamma$-ray source.

\section{Energy Reconstruction}

Stereoscopic observations of atmospheric showers with four VERITAS telescopes
enable accurate localization of the shower axis in the ground plane. Thus the impact
distances from the shower axis to the system telescopes can be calculated in a straightforward manner.
The generic reconstruction algorithm \citep{Konopelkoetal1999} is based on a simultaneous
use of image orientation in several telescopes for each individual event. The accuracy of
such reconstruction is limited by the uncertainties in the determination of the image orientation.
By observing $\gamma$-ray showers at zenith angles less than $45^\circ$ and restricting the
impact distances to less than 250~m, the average accuracy in evaluation of the telescope impacts
is better than 10~m. If the distance from the shower axis to the telescope ($r_i, i=1,n$, where
$n$ is the number of recorded images) is known, the primary energy of the air shower can
be evaluated using the inverse function of the image size with respect to the shower energy
$E_i = F(S_i, r_i,\theta)$ \citep{Konopelkoetal1999}. Here $S_i$ stands for the image size (the total
number of photoelectrons in the image), $r_i$ is the impact distance, and $\theta$ is the zenith
angle. This function can be well represented by a multi-variable look-up table, which
contains the mean energy for Monte Carlo simulations across the range of image sizes
and impact distances of recorded $\gamma$-ray showers. Such look-up tables were created
for a number of zenith angles. Finally, the shower energy can be computed by averaging
over all reconstructed energies for individual telescopes $E_i, i = 1,n$, as
$E_o = \sum_i w_i E_i$, where $w_i$ is the statistical weight ($\sum_i w_i = 1$). Rather
accurate and robust estimations can be achieved for $w_i = 1/n$. The energy resolution of
the VERITAS array of four imaging air Cherenkov telescopes averaged over the entire dynamic energy range is 15-20\%.
Note that the energy resolution is unavoidably limited by the fluctuations in image size
for a given shower energy.

In order to control any possible systematic biases in the energy reconstruction, one can
use the error in the reconstructed energy, $\delta E=(E-E_o)/E_o$ as a function of the true
energy, $E_o$. Even though this error usually does not exceed a 5\% level over the
energy range from 150~GeV to 10 TeV, a positive bias can be observed at energies close to
the threshold and a negative drop-off can be seen at very high energies. These biases are an
intrinsic feature of the reconstruction algorithms and have been well understood using detailed
Monte Carlo simulations. To diminish any noticeable effect of these biases on the measured
energy spectrum, one can limit the lowest and highest energy by requiring that the energy bias
does not exceed 20\% (see, e.g., \citet{aharCrab}). Note that the effective energy
range chosen for the spectrum evaluation substantially depends on the zenith angle of
observations as well as on the actual setup of a system configuration.

\section{Spectral Analysis}

Despite the fact that stereoscopic observations with four VERITAS telescopes provide very efficient
rejection of the cosmic ray background, the sample of selected $\gamma$-ray-like events
still contains a substantial fraction of background cosmic rays. In order to remove any effect
of background on the reconstructed energy spectrum of $\gamma$ rays, a similar energy
reconstruction procedure has to be applied to the events acquired from a number of purely
background regions as defined in the wobble-mode analysis. This enables a proper estimate of the
background contamination, which has to be subtracted from the signal region. The resulting energy-dependent $\gamma$-ray
rate can be used for the spectrum evaluation by applying a specific response matrix, which
handles various zenith angles, system configurations, observational modes, analysis setups,
etc. Such a response matrix represents a complete set of effective collection areas of the
instrument, which can be derived using detailed Monte Carlo simulations. The CORSIKA
shower simulation code (see \citet{Maier2007}) was used to generate the $\gamma$-ray- and
cosmic-ray-induced air showers over the accessible range of zenith angles and in the
energy range between 50 GeV and 100 TeV, assuming the $\gamma$-ray energy
spectrum to be a power law with an index of 2.0. Simulations of the VERITAS response were carried out
using the {\it GrISU} code, developed by the Grinnell College and Iowa State University groups (e.g., see \citet{Maier2007}).
Simulations were compared with data in great detail.
In order to avoid any remaining small energy biases in the energy reconstruction
discussed above, it was necessary to compute the effective collection areas as a function of the
reconstructed energy rather than true shower energy (see, e.g., \citet{aharCrab}). This helps to
complete the unfolding of the intrinsic source spectrum.

\begin{figure}[btp]
\includegraphics [width=1.0\textwidth]{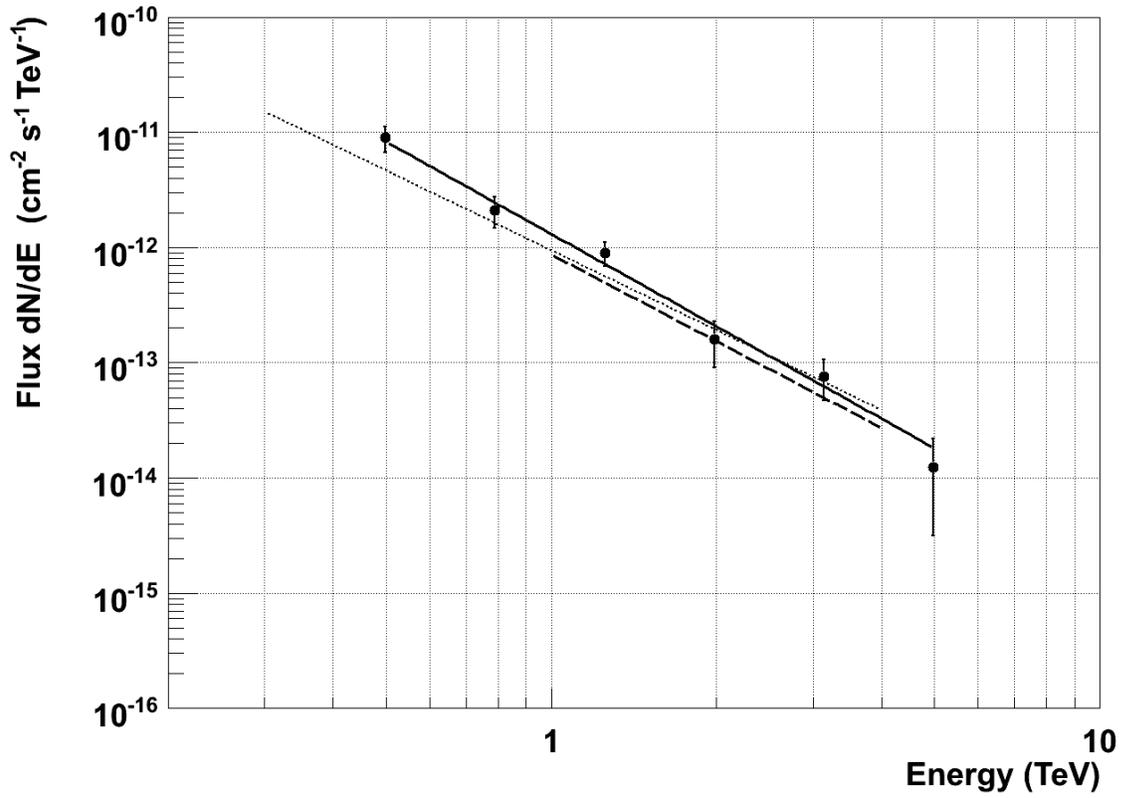}
\caption{Differential energy spectrum of TeV $\gamma$ rays from Cassiopeia~A, measured with VERITAS. Also shown are the HEGRA (dashed line) and the MAGIC (dotted line) energy spectra adapted from \citep{HEGRA2000} and \citep{Albertetal2007}, respectively.}
\label{fg4}
\end{figure}

The complete spectrum evaluation procedure has been tested on the Crab Nebula data taken during 2007/2008
observation season. 
%The power-law fit to the data gives
%\begin{equation}
%\frac{dN_\gamma}{dE}=(3.73\pm0.30)\times 10^{-11} (E/1\,TeV)^{-2.68\pm0.13} \, \rm cm^{-2}s^{-1}TeV^{-1},
%\label{crab}
%\end{equation}
%with the flux normalization at 1~TeV as
%\begin{equation}
%F_\gamma(>1\,TeV)=(2.21\pm0.24)\times 10^{-11}\, \rm cm^{-2}s^{-1}.
%\end{equation}
The result is consistent with the previous measurements of the Crab Nebula spectrum with
HEGRA~\citep{crabhegra04}, HESS~\citep{aharCrab}, and MAGIC~\citep{crabmagic08}, as well as measurements made using VERITAS data but different analysis tools.
Spectral analysis of the Cassiopeia~A data leads to the energy spectrum
\begin{equation}
\frac{dN_\gamma}{dE}=(1.26\pm0.18)\times 10^{-12} (E/1\,TeV)^{-2.61\pm 0.24_{stat} \pm 0.2_{sys}} \, \rm cm^{-2}s^{-1}TeV^{-1},
\end{equation}
with the flux normalization
\begin{equation}
F_\gamma(>1\,TeV)=(7.76\pm0.11)\times 10^{-13}\, \rm cm^{-2}s^{-1}.
\end{equation}
A power-law fit to the Cassiopeia~A spectrum  yields a $\chi^2 = 2.15$ for 4 degrees of freedom with a chance probability of $P=0.71\%$.   
This result is in good agreement with the HEGRA spectrum estimate \citep{HEGRA2000} as well as the recently published MAGIC spectrum \citep{Albertetal2007} (see Figure~\ref{fg4}). Currently, the rather limited exposure on Cassiopeia~A with VERITAS limits spectral measurements to the energy range from 300~GeV to 5~TeV. The  spectrum measured over this interval does not reveal any break of a spectral slope or cutoff. Limiting the upper energy bound for the spectral fit at 3~TeV results in a flatter spectrum index ($\Gamma$=2.4), even though the combination of the statistical and systematic errors does not allow us to draw any firm conclusion. Despite the fact that no $\gamma$-ray events were detected above 6~TeV, the statistical significance of this result is not sufficient for any firm statement regarding possible deviation of the spectral shape from a simple power law. Any other, more sophisticated, multi-parameter fit functions are not favored over a simple power law. Adding to the spectral fit an exponential cutoff term, $dN_\gamma/dE \propto E^{-\Gamma}e^{-E/E_o}$, does not improve the result but rather degrades the quality of the fit. For the cutoff energy $E_o=2$~TeV ($\Gamma=2.35$)  the fit gives the $\chi^2$-value of 3.1 ($P=0.54\%$). Further reduction of the cutoff energy down to $E_o$=1~TeV  ($\Gamma=1.46$) destroys the spectral fit ($\chi^2$=5.4, $P=0.25\%$). This result shows that  the value of possible high-energy cutoff in the Cassiopeia~A spectrum is outside the energy range measured here.
%to be far beyond a multi-TeV energy range.      
Future observations with VERITAS will help to extend the spectral measurements for Cassiopeia~A SNR.

\section{Discussion}

Understanding the mechanism of particle acceleration in isolated SNR shocks is of great interest.
The question of whether or not the very-high-energy $\gamma$--ray emission of Galactic
supernova remnants implies
a sufficiently high flux of charged cosmic rays, merging into
a steady flux of Galactic cosmic rays, remains one of the most stimulating rationales for
ground-based $\gamma$--ray astronomy. 
Following initial, simplified estimates of the expected
$\gamma$-ray flux from Galactic SNR \citep{Druryetal1994}, more refined models have been developed to
describe particle acceleration in, and high-energy emission from, Cassiopeia~A
\citep[e.g.][]{atoyan2,Berezhko2003}.
Cassiopeia~A is one of the best-studied SNR in the Galaxy, and
a multitude of observations in different wavebands constrain its physical properties
and hence the environment in which particle-acceleration processes operate.
The analysis of scattered optical light indicates that 
Cassiopeia A was of type IIb and originated from the collapse of the helium core of a red supergiant that had lost most of its hydrogen envelope before exploding \citep{krauseetal2008}. 

Cassiopeia~A is a very bright radio \citep{bell1975} and X-ray source \citep{holt94}.
\citet{bork96} have modeled the thermal X-ray emission, including the present size and expansion
rate. They concluded that the fast wind of the final blue-supergiant stage of the progenitor has
swept into a dense shell the wind material from the earlier red-supergiant phase.
The SNR blast wave has already passed through and accelerated the dense
($n_H\simeq 15\ {\rm cm^{-3}}$) shell of circumstellar material (CSM). About $8\ {\rm M_\odot}$
of X-ray-emitting swept-up gas is found in Cassiopeia~A, mostly in the form of heavy elements and predominantly located in the CSM shell and the outer, unperturbed, red-supergiant wind
\citep{willingaleetal.2003}. The presence of a jet, numerous slow-moving flocculi, and the 
general asymmetry of the remnant requires careful 3-D modeling of the supernova explosion 
\citep{laming+wang2003,youngetal2006}, which has recently been used to demonstrate the absence of a 
WR phase of the progenitor \cite{vanveelen2009}.

The high gas density combined with the high radio flux observed from Cassiopeia~A
permit an estimate of the
magnetic field strength, because the radio-emitting electrons must produce non-thermal bremsstrahlung
between 100~MeV and 10~GeV \citep{cowsik}. An upper limit on the GeV-band flux has been derived
using EGRET data \citep{Espositoetal.1996}, which in a one-zone model leads to a lower limit on the
magnetic field, $B\gtrsim 0.4$~mG \citep{atoyan2}. A high magnetic-field strength strongly
limits the flux of an inverse-Compton emission component in the TeV band on account of
its direct relation to synchrotron X-ray emission \citep{pohl96}, which is observed up to 120~keV
\citep{Allenetal.1997,Favataetal.1997}. The observed hard X-ray emission also includes lines
from the $^{44}$Ti decay chain \citep{Renaud06}. 

The non-thermal X-ray emission predominantly originates from filaments and knots in the
reverse-shock region of Cassiopeia~A \citep{HelderVink08},
some of which are variable in flux on timescales of years
\citep{uchiyama08}. Both the filaments themselves and their flux variability require a strong magnetic
field, but estimates of its exact amplitude depend on their detailed interpretation. Typically, one
obtains somewhat higher values for the magnetic-field strength, if one assumes
the size and variability timescale are determined by electron energy losses
\citep{VinkLaming2003,uchiyama07,uchiyama08}, as opposed to scenarios involving the rapid damping of a turbulently
amplified field \citep{pyl05} or localized spikes in dynamical magnetic turbulence \citep{bykov}.
The complicated structure of Cassiopeia~A, including fast-moving clumps of ejecta and knots of high radio
brightness, has triggered the development of many scenarios involving first- and second-order
Fermi-type acceleration at various locations \citep[e.g.][]{scott75,jones94,atoyan1}.

The presence of a large flux of high-energy electrons in the reverse-shock region, responsible for the
non-thermal radio-to-X-ray emission, will also produce high-energy $\gamma$-ray emission through
non-thermal bremsstrahlung and inverse-Compton scattering \citep[e.g.][]{atoyan2}.
Based on that leptonic
emission, Cassiopeia~A would appear in VERITAS data as a disk- or ring-like source with outer radius
$R_l\lesssim 2'$ \citep{uchiyama08}. If, on the other hand, the VHE $\gamma$-ray emission from Cassiopeia~A
were dominated by $\pi^o$-decay photons produced in inelastic collisions of relativistic protons, the
location of the particle-acceleration sites is less constrained by data in other wavebands, and
substantial acceleration of cosmic-ray protons could proceed at the outer blast wave
\citep{Berezhko2003}. The size of Cassiopeia~A in VERITAS data could therefore be slightly larger than for
leptonic scenarios, with an outer radius $R_h\lesssim 2.5'$. However, both predicted angular extensions are substantially 
less than the current angular resolution of VERITAS.  The VERITAS data are consistent with a point-like $\gamma$-ray source.
Since the extended TeV $\gamma$-ray emission has not been resolved with VERITAS yet, current results remain
fully consistent with the hypothesis that the VHE $\gamma$-ray emission originates from the central part of
Cassiopeia A, where a compact object has been observed at longer wavelengths.
Nevertheless, we encourage modeling the spatial distribution of VHE $\gamma$-ray emission from Cassiopeia~A
in preparation for the next
generation of imaging atmospheric Cherenkov telescopes,
CTA\footnote{\url{www.cta-observatory.org}} and
AGIS\footnote{\url{www.agis-observatory.org}}, both of which are projected to
have an angular resolution better than that of VERITAS by a factor of at least 2.

None of the published calculations of VHE $\gamma$-ray production correctly predict both the flux and the
spectrum observed with VERITAS. \citet{atoyan1,atoyan2} have carefully modelled the acceleration,
propagation, and photon-emission spectra of high-energy electrons. For the parameters chosen for their
displayed $\gamma$-ray spectra, the TeV-band emission is a mixture of non-thermal bremsstrahlung and
inverse-Compton scattering that would account for about 25\% of the flux observed with VERITAS and
feature a softer spectrum than observed ($\propto E_\gamma^{-3.2}$). We can speculate that a better fit
may be achieved by a small reduction of the magnetic-field strength and/or an increase
in the cut-off energy of the electron injection spectrum, which astrophysically is determined by the
details of the acceleration process, the magnetic-field strength in the acceleration region, and
efficiency limitations imposed by the geometry of the acceleration region. However, a low GeV-band
flux measurement or upper limit derived with Fermi-LAT would imply a magnetic field stronger than that
assumed by \citet{atoyan2}.

Models of hadronic VHE $\gamma$-ray emission are somewhat less constrained by radio and X-ray
data than are their leptonic counterparts. An indirect relation exists in that a very efficient
acceleration of cosmic-ray nuclei by shock fronts leads to a modification of such shocks,
resulting in relatively soft spectra below a GeV particle energy and rather hard spectra around a TeV
\citep[e.g.][]{blandford87,berezhko99}, although a cosmic-ray-induced strong magnetic field can
substantially reduce the shock modification compared with the naive unmagnetized case
\citep{caprioli08}. The soft
radio spectrum observed from Cassiopeia~A is indeed consistent with
non-linear kinetic models of cosmic-ray acceleration in SNR \citep[e.g.][]{Berezhko2003}.
However, the hard spectra
predicted beyond a particle energy of 1~TeV have not been observed to date. In fact, the VHE $\gamma$-ray
spectra measured from shell-type SNR are all well described by either a power law with photon index
$s\lesssim -2.2$, or a power law with gradual roll-off \citep[e.g.][]{huang07}; this suggests
that the acceleration of cosmic-ray protons beyond 1~TeV must be less efficient than previously
thought, if the observed VHE $\gamma$-ray emission completely arises from interactions of these protons.

Except for the electron energy losses, the high-energy cut-off in the spectrum of accelerated
protons is determined by the same
physical processes as that of the electrons, the details of which are not well understood.
The published models may therefore be too optimistic in predicting the cut-off energy
\citep[e.g.][]{Berezhko2003}. In any case, more work is required to better understand the high-energy
end of cosmic-ray proton spectra accelerated in SNR \citep[e.g.][]{ellison08}.
In addition, the role stochastic
particle acceleration plays in SNR needs to be explored in more detail \citep{liu08}. Observationally,
the next step toward a better understanding of particle acceleration in Cassiopeia~A will be measuring
the GeV-band $\gamma$-ray spectrum with the Fermi-LAT.

\section*{Acknowledgements}

\noindent
This research was supported by grants from the U.S. Department of Energy, the U.S. National
Science Foundation and the Smithsonian Institution, by NSERC in Canada, by Science Foundation
Ireland and by STFC in the UK. The VERITAS collaboration acknowledges the NASA support on the {\it Fermi}
GST LAT Grant \#NNX08AV62G.

\pagebreak

%%%% Table 1. %%%%%%%%%%%%%%%%%%%%%%
\begin{table}[!t]
\centering
\caption{Summary of analysis cuts.}
\vspace*{2mm}
\begin{tabular}{lllll}\hline
Set & Flux (Crab) & MSW $(^\circ)$ & MSL $(^\circ)$ & $\theta\, (^\circ)$ \\ \hline
A & 0.03 & [0.05,1.08] & [0.05,1.19] & 0.13 \\
B & 1 & [0.05,1.1] & [0.05,1.39] & 0.158 \\ \hline
\end{tabular}
\label{t1}
\end{table}

%%%% Table 2. %%%%%%%%%%%%%%%%%%%%%%
\begin{table}[!t]
\centering
\caption{Results of data analysis.}
\vspace*{2mm}
\begin{tabular}{lllll}\hline
Source: & Crab Nebula & & Cas A &  \\ \hline \hline
Exposure (hr) & 3.0 & & 21.8 & \\
Set of cuts & A & B & A & B \\
On events & 891 & 1298 & 625 & 1277 \\
Off events$^*$ & 480 & 841 & 3538 & 6164 \\
Significance ($\sigma$) & 47.6 & 50.5 & 8.3 & 7.0 \\
$R_\gamma\, \rm (min^{-1})$ & 5.07$\pm$0.18 & 7.10$\pm$0.22 & 0.148$\pm$0.019 & 0.191$\pm$0.028 \\
$R_{CR}\, \rm (min^{-1})$ & 0.35 & 0.80 & 0.32 & 0.77 \\ \hline
%$R_{CR}/R^2_\gamma \rm (min^{-1/2})$ & 0.0137 & 0.0158 & 14.7 & 21.06 \\ \hline
\end{tabular}

\vspace*{2mm}
$^*$A total number of Off events was accumulated over a few similar circular regions.
\label{t2}
\end{table}

\pagebreak

%%%%% Figure 1 %%%%%%%%%%%%%%%%%%%%%%%%%%%%%%%%
%\begin{figure}[btp]
%\includegraphics [width=1.0\textwidth]{figure1}
%\caption{Scaled parameters of transverse (upper panel) and longitudinal (lower panel) angular size of the
%Cherenkov light images as a function of primary energy of the $\gamma$-ray-induced air showers. The dashed
%lines indicate corresponding analysis cuts.}
%\label{fg1}
%\end{figure}

%%%%% Figure 1 %%%%%%%%%%%%%%%%%%%%%%%%%%%%%%%%

%%%%% Figure 4 %%%%%%%%%%%%%%%%%%%%%%%%%%%%%%%%

\end{document}